\documentclass[pra,
superscriptaddress,
nofootinbib,msmath,amssymb,aps,
floatfix,twocolumn]{revtex4-2}
\usepackage{graphicx}
\usepackage{subcaption}
\usepackage{amsmath,amssymb,xcolor,dsfont}
\usepackage{url}
\usepackage[colorlinks=true, linkcolor=blue, citecolor=red, urlcolor=cyan]{hyperref}
\usepackage{tikz}
\usepackage{quantikz}
\usepackage[capitalize]{cleveref}
\usepackage{diagbox}

\graphicspath{{./figures/}}

\newcommand{\figscale}{0.46}
\newcommand{\figscaletwo}{0.51}

\begin{document}

\title{Towards Quantum Advantage in Sparsified Bosonic SYK Models}

\author{Vaibhav Gautam}
\affiliation{Kavli IPMU (WPI), UTIAS, The University of Tokyo, Kashiwa, Chiba 277-8583, Japan}
\author{Atsushi Matsuo}
\affiliation{IBM Quantum, IBM Research---Tokyo, Tokyo 103-8510, Japan}
\author{Masahito Yamazaki}
\affiliation{Kavli IPMU (WPI), UTIAS, The University of Tokyo, Kashiwa, Chiba 277-8583, Japan}
\affiliation{Graduate School of Physics, University of Tokyo, Tokyo 113-0033, Japan}
\affiliation{Trans-Scale Quantum Science Institute, 
    University of Tokyo, Tokyo 113-0033, Japan}
\affiliation{Center for Data-Driven Discovery (CD3), Kavli IPMU, University of Tokyo, Chiba 277-8583, Japan}

\begin{abstract}
We advocate the sparsification of bosonic SYK models as a promising arena for the exploration of quantum advantage.
We initiate the study of quantum simulations of the models, both in classical simulators and on quantum devices implemented using superconducting qubits. 
We point out subtleties in the quantum simulations of highly chaotic systems, which should be addressed in the future search for quantum advantage.
\end{abstract}

\maketitle

\section{Introduction and Summary}

\subsection{Quantum Advantage and Quantum Chaos}

Over the past few years, we have witnessed remarkable advances in quantum technologies, and we are now at a juncture where large-scale quantum computation has become a realistic and actively pursued goal.

One of the most important questions in quantum computation is to identify and characterize problems for which quantum computers outperform their classical counterparts \cite{Feynman:1981tf, Shor:1994jg, Deutsch:1985vkq, Preskill:2018jim}. Achieving such a ``quantum advantage''\footnote{See \cite{Lanes:2025kzh} for a more concrete definition of quantum advantage.} will undoubtedly be one of the major milestones in the development of quantum computing.

Since its inception \cite{Feynman:1981tf}, quantum simulations of physical quantum-mechanical systems have been proposed as one of the most promising arenas for exploring quantum advantage---the system is intrinsically quantum mechanical, and the complexity of the physical states is expected to grow rapidly due to the Hamiltonian time evolution. The latter is particularly pronounced for systems with \emph{quantum chaos},
where the tiny variations of the initial states are quickly amplified, signaling the difficulty of classical simulations. It is therefore natural to look for ``the most chaotic quantum mechanical system
in the Universe''
as a venue for quantum advantage. It turns out that the Sachdev-Ye-Kitaev (SYK) model \cite{Sachdev:1992fk,Kitaev:2015,Maldacena:2016hyu} is 
precisely such a system.

To formulate this statement more quantitatively,
let us introduce an out-of-time-ordered correlator (OTOC) \cite{larkin1969quasiclassical,Shenker:2013pqa} of two 
local operators $V$ and $W$, inserted at times $0$ and $t$:
\begin{align}
C(t) :=\langle W^{\dagger}(t) V^{\dagger}(0) W(t) V(0) \rangle 
\end{align}
Note that this correlation function involves operators inserted at times $0, t, 0, t$, thereby involving both forward and backward time evolutions (and hence the name ``out-of-time-ordered'').
For a generic operator $W$ and the Hamiltonian $H$, the complexity of the time-evolved operator
$W(t) = U(t)^{\dagger}\, W\, U(t)$ (with $U(t):= e^{-i Ht}$
being the time-evolution operator) will grow exponentially, 
signaling the difficulty of classical simulations.
On the other hand, quantum simulations for measurement of OTOC have been progressing at a fast pace in the past decade, starting with experimental quantum simulations \cite{Li:2016xhw,Garttner:2016mqj} to studying operator spreading and information scrambling \cite{Landsman:2018jpm,Vermersch:2018sru, Blok:2020may, Mi:2021gdf, Braumuller:2021cic, Geller:2021kpx,Seki:2024rfx}.

The measurement of the time evolution of OTOC is a promising task for exploring possible quantum advantage.
In particular, the exponential growth of the OTOC in early times
defines the Lyapunov exponent $\lambda_L$:
\begin{align}
C_{V, W}(t) \sim C_1  + C_2 \, e^{\lambda_L t},
\end{align}
and this exponent is one of the proposed quantitative characterizations of quantum chaos \cite{aleiner1996divergence, aleiner1997role, agam2000shot}. Note that this definition refers to the early-time growth of the OTOC---while quantum chaos can also be discussed by late-time behaviors of OTOC, the early-time behavior is particularly 
suitable for demonstrating quantum advantage, since measurements in shorter time scales involve a smaller number of time steps and thus are less susceptible to noise.

The exponent satisfies the general bound \cite{Maldacena:2015waa}
\begin{align}
    \lambda_L \le \frac{2\pi k_B T}{\hbar} ,
\end{align} where $k_B$ is the Boltzmann constant, 
$T$ is the temperature of the system. 
The bound involves the Planck constant $\hbar$ and hence is an intrinsically quantum bound.
The SYK system is special in that it saturates this bound; relatedly, we expect the system to be dual to a weakly-coupled holographic gravity dual.
  
\subsection{Bosonization and Sparsification}

Given the importance of the SYK model in both quantum chaos and holography, there have been many attempts, either theoretical or in quantum hardware, for quantum simulations of the SYK models \cite{luo2019quantum, Baumgartner:2024ysk, asaduzzaman2024sachdev}. It turns out, however, to be challenging to simulate the SYK model on current quantum simulators. For example, the model involves all-to-all entanglements of qubits and requires many SWAP gates in localized architectures,
such as superconducting qubits. 
While it is worthwhile to continue these lines of exploration (and indeed the problem may well be resolved in architectures with all-to-all connectivity, such as ion traps \cite{granet2025simulating}),
it is fair to say that we are still far from a realistic study of the SYK model as an arena for quantum advantage.

In this paper, we propose to study simplified models
which dramatically simplifies the computational cost of the system, while making the system more tractable in present-day computers. 
Our discussion involves two ingredients, which are conceptually independent of each other:

\paragraph{Bosonification:}
First, we replace the fermions of the SYK model by bosons, so that we have the bosonic version of the SYK model introduced in Refs.~\cite{Swingle:2023nvv,Hanada:2023rkf}.
We call this the \emph{bosonification}, to distinguish this from the bosonization 
concerning the duality transformation.

While much remains to be studied for this bosonic SYK model, 
the model was solved in the large $N$ limit and shows a behavior similar to its fermionic counterpart, if not maximally chaotic.
This suggests that the bosonic SYK model will still be a highly chaotic system and thus a promising system to demonstrate quantum advantage.

\paragraph{Sparsification:}
Second, we will discuss the sparsification of the model,
where sparsification involves choosing individual Pauli operators of the Hamiltonian 
with a probability $p$. In practice, we choose $p\sim \mathcal{O}(N^{-3})$
which dramatically simplifies the model in the large $N$ limit.
The sparsification of the fermionic SYK model was proposed in Ref.~\cite{Xu:2020shn}
and studied further in Refs.~\cite{Garcia-Garcia:2020cdo, Orman:2024mpw, Tezuka:2022mrr}. It was found that there is a transition from the chaotic regime to the non-chaotic regime as we change the probability $p$ below a threshold value.

In this paper, we advocate for this simplified model for quantum advantage by 
\emph{combining the two ingredients}---the  sparsification of the bosonic SYK model,
which we call the \emph{sparsified bosonic SYK model}. We believe this is among the simplest models that is simultaneously 
practical for quantum simulations and 
preserves the essence of the physics of the original SYK model.

\subsection{Summary of the Results}

We initiate a study of OTOC's for  
the unsparsified and sparsified bosonic SYK models in both classical simulators and 
quantum hardware. 

The goal of this paper is two-fold. 

First, we study the physical effect of sparsification by varying the sparsification probability $p$.
We find a transition from chaotic to non-chaotic behavior as we lower the value of the $O(N^0)$ coefficient $\kappa$.

Second, we report on a preliminary study on the
quantum simulations on the IBM superconducting devices.
While this is conceptually simple and 
we follow the standard Trotterization scheme,
we find important subtleties in the 
present-day noisy hardware specific to our model at hand.

The rest of this paper is organized as follows. In \cref{sec:model}
we first introduce the bosonic SYK model, as well as its new sparsified variant. In \cref{sec:simulation} we summarize our methods for quantum simulations. Our results are summarized in \cref{sec:result}.
We conclude this paper in section \cref{sec:conclusion}
with comments on future directions.

{\it \bf Note added: } While this work was in progress, we became aware of a related paper studying Out-of-Time-Ordered Correlators (OTOCs) \cite{Abanin:2025rbz}.
Although related, our approach differs in that our quantum circuits arise directly from controlled simplifications of the SYK model and are therefore more firmly grounded in its underlying physics. Correspondingly, our Trotterized time evolutions are substantially more involved than those studied in Ref.~\cite{Abanin:2025rbz}. Most importantly, as discussed in the main text, the circuit depths used here are significantly larger. 
It would be valuable to pursue a more detailed comparisons between our results and those of Ref.~\cite{Abanin:2025rbz}, and to 
more fully integrate perspectives on quantum advantage, quantum chaos, and holography.

\section{Bosonic SYK model: with and without Sparsification}
\label{sec:model}

\subsection{Bosonic SYK model}

The Hamiltonian of the original SYK model \cite{Sachdev:1992fk}, which we call the fermionic SYK model in this paper, is given by the following four-Fermi interactions:
\begin{align} \label{Eq: Ham_fermionic}
    \mathcal{H}_{\text{f-SYK}} = \sqrt{\frac{6}{N^3}} \sum_{1\leq i < j < k < l \leq N} J_{ijkl}\:  \hat{\psi_i} \hat{\psi_j} \hat{\psi}_k \hat{\psi}_l
\end{align}
where $\psi_i$ ($i=1, \dots, N$) are 
$N$ copies of Majorana fermions satisfying the anti-commutation relations
$\{\hat{\psi}_i, \hat{\psi}_j \} = \mathsf{i} \delta_{ij}$,
and the coefficients $J_{ijkl}$ are sampled classically from 
the Gaussian distribution
with average and variance given by
\begin{align}
    \langle J_{ijkl} \rangle = 0, 
    \quad \langle J_{ijkl}^2 \rangle = \frac{6 J^2}{N^3}.
\end{align}
Note that this model can be regarded as a classical-quantum hybrid system, in that quantum time evolution is combined with classical ensemble average.

When we simulate this in a quantum device, we often invoke the Jordan-Wigner transformation from fermions into quantum spins, which is a non-local transformation and hence involves many circuits.
The issue is particularly severe in this case, since the model involves all-to-all connectivities between the fermions.

To alleviate this problem, in this paper, we consider the bosonic variant of the SYK model \cite{Swingle:2023nvv,Hanada:2023rkf} whose Hamiltonian is given by the following four-body interaction of \emph{bosons}:
\begin{align} \label{Eq: Ham_bsyk}
    \mathcal{H}_{\text{b-SYK}} = \sqrt{\frac{6}{N^3}} \sum_{1\leq i < j < k < l \leq N} J_{ijkl}\: \mathsf{i}^{\eta_{ijkl}} \hat{\phi}_i \hat{\phi}_j \hat{\phi}_k \hat{\phi}_l
\end{align}
where the bosons $\hat{\phi}_i$ satisfy the canonical commutation relations
$[\hat{\phi}_i, \hat{\phi}_j ]= \mathsf{i} \delta_{ij}$, and $J_{ijkl}$ are random coupling constants 
obtained from the normal distribution as before. 
The factor of $\mathsf{i}^{\eta_{ijkl}}$, where $\eta_{ijkl}$ is the number of $\hat{\sigma}_z$ in the simplified product $\hat{\phi}_i \hat{\phi}_j \hat{\phi}_k \hat{\phi}_l$, is included to ensure that the Hamiltonian is Hermitian.

Note that the idea of replacing the fermions by bosons has been discussed in other contexts in condensed matter physics, e.g.\ in the context of Bose-Hubbard models. Such a bosonification could also be natural experimentally, since this would involve replacing the fermionic species by their bosonic counterparts, as in the realizations in cold atoms, for example.

\subsection{Sparse Bosonic SYK}
Although the circuit-depth requirements for the bosonic SYK model are reduced compared to the standard SYK model, they still exceed the depth limits of current quantum hardware for reliable computation of out-of-time-ordered correlators (OTOCs), particularly at larger values of $N$. This limitation motivates the development of alternative models that retain the essential features of the theory’s chaotic dynamics while reducing circuit complexity. One possible approach is to introduce sparsity by randomly setting the coefficients of certain terms in the Hamiltonian Eq.~\ref{Eq: Ham_bsyk} to zero, thereby reducing the effective parameter space. For the SYK model, such sparsification has been shown to successfully reproduce the chaotic properties of the full theory \cite{Xu:2020shn}.

As is the case for the SYK model, we can define a sparse version for the bosonic-SYK model described above by randomly setting some $J_{ijkl}$ to zero. The Hamiltonian for the sparse bosonic-SYK model can be written as 
\begin{align} \label{Eq: sparse Ham_bsyk}
    \mathcal{H}_{\text{sparse}} = \sqrt{\frac{6}{pN^3}} \sum_{1\leq i < j < k < l \leq N} p_{ijkl} J_{ijkl}\: i^{\eta_{ijkl}} \hat{\phi}_i \hat{\phi}_j \hat{\phi}_k \hat{\phi}_l
\end{align}
where $\hat{\phi}_i$ are as described in Eq.~\ref{eq: OurMajoranas} and $p_{ijkl} \in \{0,1\}$ are randomly sampled independently with a probability $p$ for $1$ ($1-p$ for $0$).
Motivated by the analysis for the sparsified fermionic SYK model, we parametrize
\begin{align} \label{Eq: sparsity parameter}
    p = \kappa \cdot N/\binom{N}{4} ,
\end{align}
with $\kappa$ being an $O(N^0)$ sparsity parameter.
Note that in the large $N$ limit we have $p\sim O(N^{-3})$, and $O(N^4)$ terms in the Hamiltonian are truncated into $O(N)$ terms. 

\section{Quantum Simulations}\label{sec:simulation}

The quantum simulations can be decomposed into three steps: initial state preparation,
time evolution, and measurement of observables.

\subsection{Trotterized Time Evolution}

For the time evolution, we can use the standard Trotterization scheme for time evolution, once the Hamiltonian is represented by Pauli operators.

It is straightforward to realize the $N$ bosons $\mathcal{O}_{i}$ in terms of spin operators
acting on $N_Q = N/2$ qubits:\footnote{We assume for simplicity here that $N$ is even; the case of $N$ odd is similar.}
\begin{align}\label{eq: OurMajoranas}
    &\hat{\phi}_{2i-1} =  \mathds{1}_2 \otimes \mathds{1}_2 \otimes \ldots \otimes \hat{\sigma}_x \otimes \ldots \otimes \mathds{1}_2 \otimes \mathds{1}_2. \nonumber \\
    &\hat{\phi}_{2i} = \mathds{1}_2 \otimes \mathds{1}_2 \otimes \ldots \otimes \hat{\sigma}_y \otimes \ldots \otimes \mathds{1}_2 \otimes \mathds{1}_2. 
\end{align}
where $\hat{\sigma}_x$, $\hat{\sigma}_y$ are the usual Pauli matrices inserted in the $i^\text{th}$ position, and we have $i=1,2,\ldots, N_Q$.
Using the above representation, the Hamiltonian can be expressed as a sum of Pauli strings of length $N_s = N/2$, where each string contains at most four non-identity Pauli operators. This structure stands in contrast to that of the original SYK model, which typically involves denser operator terms.

Consequently, the circuit depth required for the Trotterization of the bosonic SYK model is significantly reduced compared to the fermionic SYK model, particularly at larger system sizes $N$, while still preserving the model's inherent chaotic dynamics (see Table \ref{Table:Depth comparison}). The reduced gate complexity makes the bosonic SYK model a promising candidate for simulating the dynamics of potentially maximally chaotic and holographic quantum systems on NISQ hardware. 

\begin{table}[htbp]
   \centering
    \begin{tabular}{|c||c|c|}
    \hline
    $N$ & Bosonic SYK & fermionic SYK\\
      \hline
      \hline
     6 & 82 & 56  \\
     8 & 497 & 475  \\
     10 & 1656 & 1972  \\
     12 & 3647 & 5212  \\
     14 & 7507 &  12486 \\
     16 & 16505 & 25394  \\
     18 & 28886 &  47774 \\
     20 & 47484 & 82330  \\
     \hline
    \end{tabular}
    \caption{
   Comparison of two-qubit circuit depth for one-Trotter step computation of OTOC using the interferometric protocol for the bosonic SYK and the fermionic SYK. 
    }
    \label{Table:Depth comparison}
\end{table}

For practical implementations on NISQ hardware, the circuit depths listed in Table \ref{Table:Depth comparison} remain relatively large. To further reduce both the circuit depth and the number of two-qubit gates, it is necessary to sparsify the Hamiltonian. As discussed in Sec.~\ref{subsec: Sparsify}, the chaotic properties of the full Bosonic SYK model can be preserved with only $O(N)$ randomly chosen nonzero terms. Consequently, the sparsified Bosonic SYK model yields significantly shallower circuits, making its simulation more feasible on current quantum devices.

\begin{table}[htbp]
   \centering
    \begin{tabular}{|c||c|c|c|c|}
    \hline
    $N$ & $\kappa = 0.1$ & $\kappa = 0.5$ & $\kappa = 1$ & $\kappa = 2$\\
      \hline
      \hline
     8 & 5 & 21 & 54 & 111\\
     10 & 3 & 37 & 113 & 199\\
     12 & 10 & 58 & 147 & 303\\
     14 & 6 & 90 & 188 & 397\\
     16 & 9 & 80 & 260 & 512\\
     18 & 10 & 115 & 277 & 571\\
     20 & 8 & 156 & 292 & 728\\
     \hline
    \end{tabular}
    \caption{
   Comparison of the two-qubit circuit depth for one-Trotter step computation of OTOC using the interferometric protocol for different values of sparseness parameter for the bosonic SYK model. The depth reported is an average of 25 realizations of the Hamiltonian.
    }
    \label{Table:Sparse Depth comparison}
\end{table}

\subsection{Interferometric protocol for OTOC}

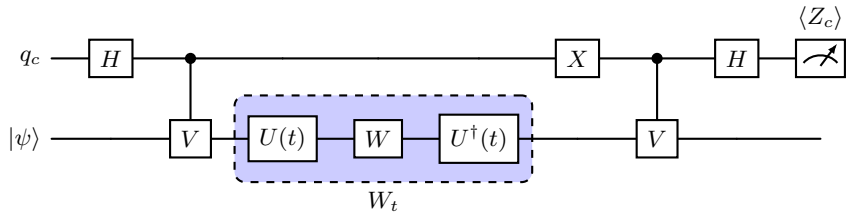
\begin{figure*}[htbp]
\begin{center}
\scalebox{1.0}{
    \begin{quantikz}
        \lstick{$q_c$} & \gate{H} & \ctrl{1} & & & & \gate{X} & \ctrl{1} & \gate{H} & \meter{\langle Z_c \rangle}\\ 
        \lstick{$\ket{\psi}$} & & \gate{V} & \gate{U(t)} \gategroup[1,steps=3,style={dashed,rounded corners,fill=blue!20, inner xsep=2pt},background,label style={label position=below,anchor=north,yshift=-0.2cm}]{{$W_t$}} & \gate{W} &  \gate{U^\dagger(t)}& &\gate{V} & &
    \end{quantikz}
}
\end{center}
\caption{Circuit implementing the interferometric protocol to compute the OTOC. The topmost wire corresponds to the control qubit $q_c$ while the lower wire corresponds to the system qubits. 
The time evolution $U(t)=e^{-i Ht}$ is implemented by Trotterization. }
    \label{fig: Interferometer Circuit}
\end{figure*}

Let us next comment on the measurement of the OTOC. The most obvious method to do this is to repeat twice both forward
and backward time evolutions. This is not ideal, however, in noisy devices, since OTOC involves detailed cancellations between backward and forward time evolutions that will
be affected by different noises.
We therefore employed the interferometric protocol \cite{Swingle:2016var,Swingle:2018xvb} depicted in \cref{fig: Interferometer Circuit}. This involves an extra ancilla qubit, which we will call the control qubit, which is eventually measured in the $X$-basis. The control qubit $q_c$ is initially prepared in the state $\ket{0}_c$ while the system qubits can be prepared in any state $\ket{\psi}$, i.e.\ the whole system in the initial state $\ket{0}_c \otimes \ket{\psi}$. 
The $\langle Z_c \rangle$ on the control wire in \cref{fig: Interferometer Circuit} indicates the measurement of the expectation value of the operator $Z \otimes \mathds{1}_N$ with Pauli-$Z$ acting on the control qubit. It can be checked that the circuit leads to the computation of the real part of $C(t)$. 

\subsection{Simulation Setups}
For all simulations, the system is prepared in the default initial state $\ket{\psi} = \ket{00...0}$.  
Our simulations are performed on the \texttt{ibm\_kawasaki} device, which employs the Heron r2 processor with 156 qubits. The median error rates and thermalization times of the QPU are listed in Table \ref{Table:error rates}.
\begin{table}[htbp]
   \centering
    \begin{tabular}{|c|c|}
    \hline
     Median readout error & 4.761e-3\\
     \hline
     Median CZ error & 1.359e-3 \\
     \hline
     Median SX error & 1.851e-4\\
     \hline
     Median T1 & 337.11 $\mu s$\\
     \hline
     Median T2 &  159.69 $\mu s$\\
     \hline
    \end{tabular}
    \caption{
   Error rates for the \texttt{ibm\_kawasaki} device at the time of simulations.
    }
    \label{Table:error rates}
\end{table}

For each simulation performed, we use 4096 shots. As one of our objectives is to study the chaotic system on a NISQ device, in this preliminary analysis we do not attempt any error mitigation and use the in-built Twirled Readout Error eXtinction (TREX) measurement twirling for suppressing readout errors.

However, we do report our results along with an attempt to `renormalize' the OTOC using the renormalization scheme introduced in Ref.~\cite{Swingle:2018xvb}. As per this scheme, let $C^{\text{noisy}}_{V,W}(t)$  be the value of the OTOC computed on a noisy device using the interferometric protocol described in \cref{fig: Interferometer Circuit} for operators $V$ and $W$. Then, the renormalized value of the OTOC can be given as:
\begin{align}\label{Eq: renormalized OTOC}
    C(t) \approx \frac{C^{\text{noisy}}_{V,W}(t)}{C^{\text{noisy}}_{V,\mathds{1}}(t)} 
\end{align}

For our purposes, the operator $V = Z_1$ i.e.\ Pauli-$\sigma_z$ acting on the first qubit ($ZII\ldots$ in Pauli string notation) and the operator $W = Z_2$ i.e.\ Pauli-$\sigma_z$ acting on the second qubit ($IZI\ldots$ in Pauli string notation). For this case, for an ideal simulation, $C^{\text{ideal}}_{V,\mathds{1}}(t) = 1$ for all $t$. 

\section{Results}\label{sec:result}

In this section, we present our preliminary results on the quantum simulations of bosonic SYK models and their sparsifications.

\subsection{Ensemble Average}
\begin{figure}[htbp]
\begin{center}
\scalebox{\figscale}{
\includegraphics{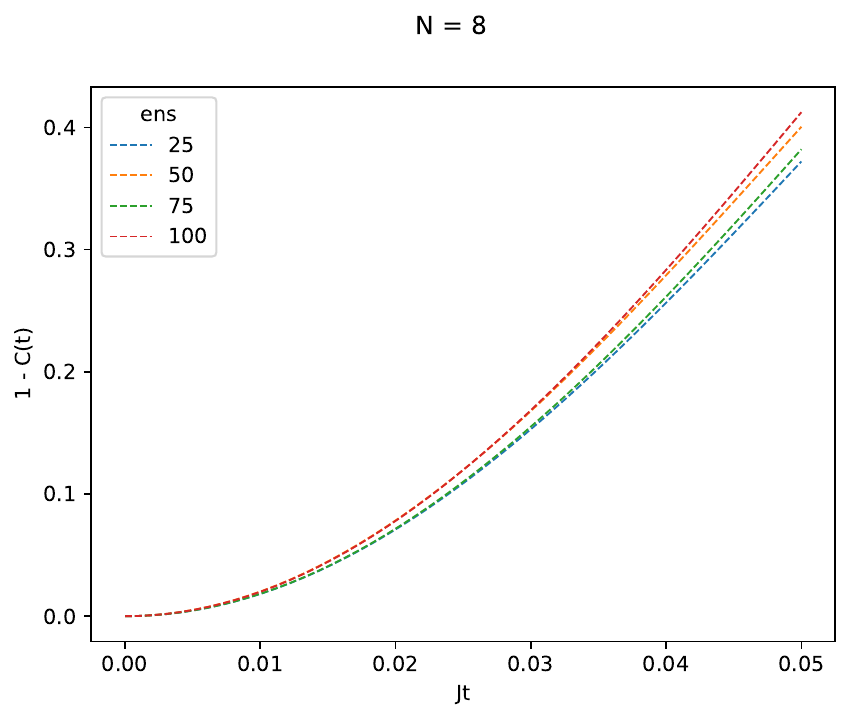}}
\scalebox{\figscale}{
\includegraphics{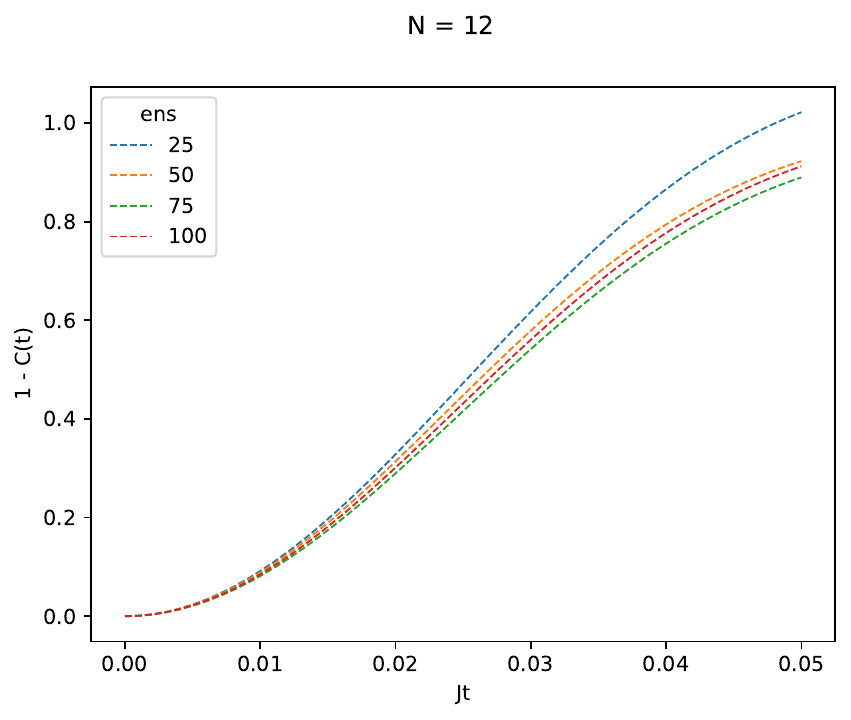}}
\scalebox{\figscale}{
\includegraphics{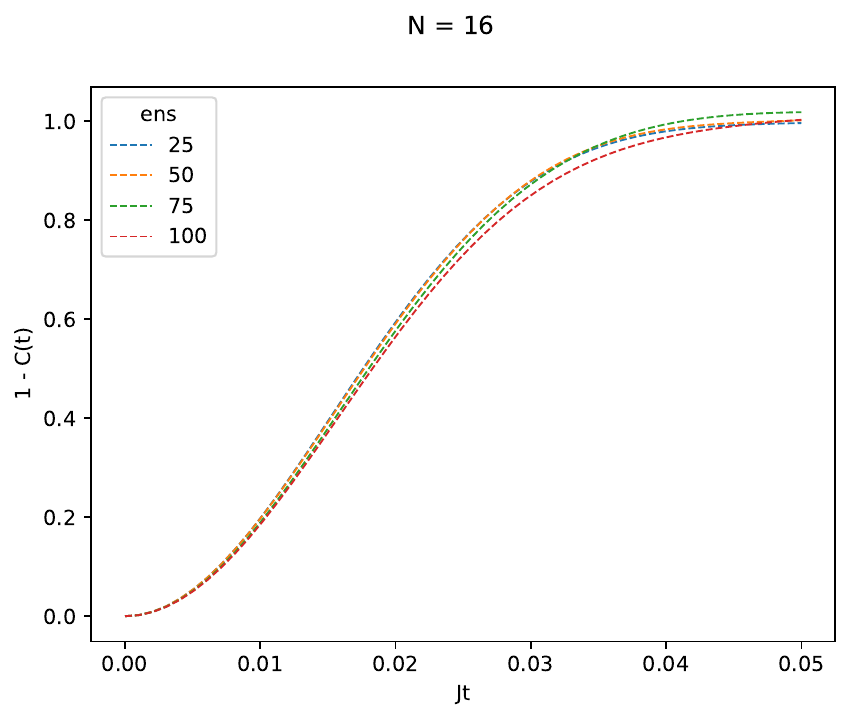}}
\scalebox{\figscale}{
\includegraphics{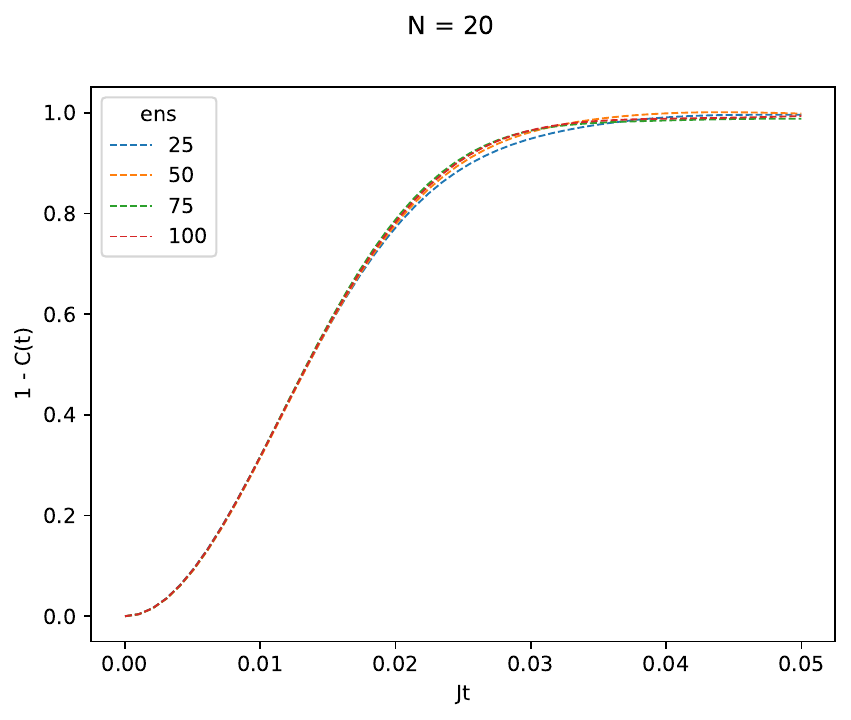}}
\end{center}
\caption{ Comparison of the average value of OTOC for different numbers of ensembles for the full bosonic SYK model.  }\label{fig: Ensemble Comparison}
\end{figure}

Since the system under consideration contains $O(N^4)$ random parameters, it is advantageous to perform an ensemble average over multiple realizations of the Hamiltonian in order to capture the typical behavior of the system. In the context of the fermionic SYK/JT gravity correspondence, this ensemble averaging corresponds to summing over different geometries in the dual gravitational theory, including various wormhole contributions. Accordingly, we compute the OTOC averaged over an ensemble of Hamiltonians. We numerically investigate the dependence of the results on the number of ensembles for the full bosonic SYK model (\cref{fig: Ensemble Comparison}). For larger values of $N$, the average value of $C(t)$ stabilizes for as few as 25 ensembles, showing little deviation.  

\subsection{Sparsification} \label{subsec: Sparsify}

\begin{figure}[htbp]
\begin{center}
\scalebox{\figscaletwo}{
\includegraphics{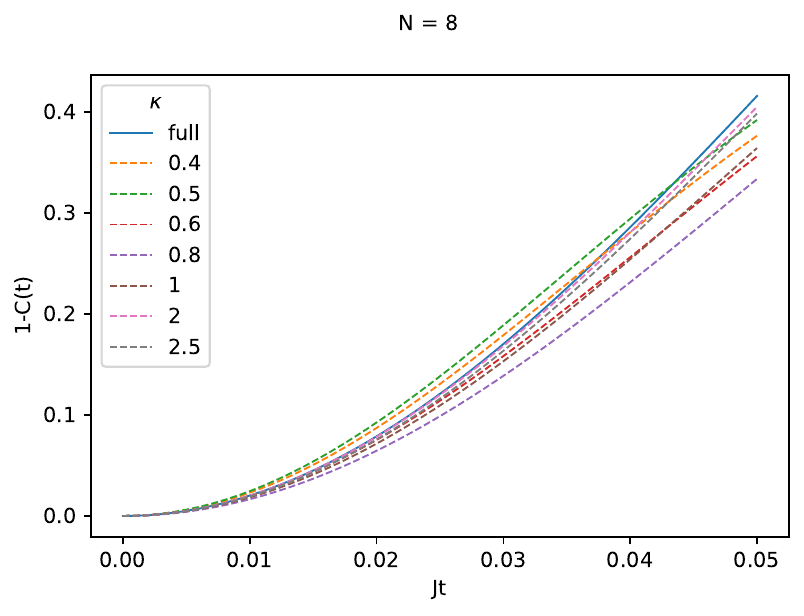}}
\scalebox{\figscaletwo}{
\includegraphics{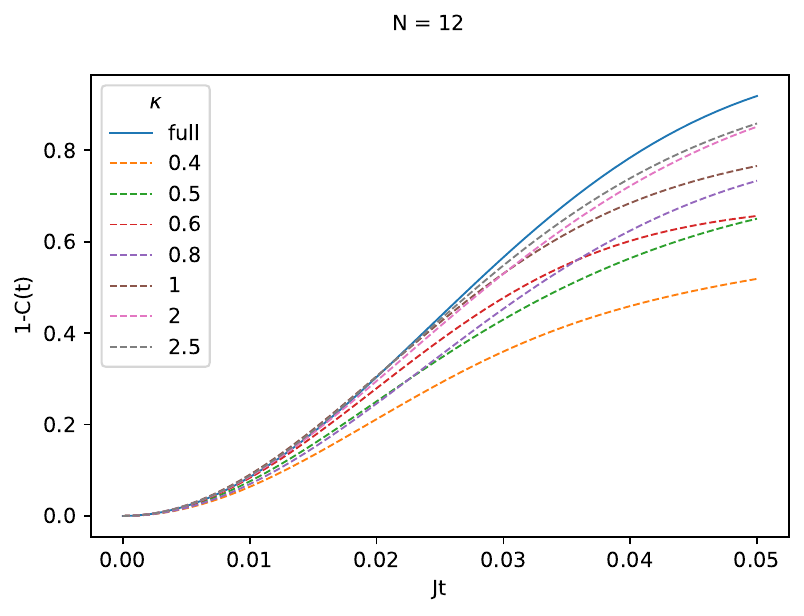}}
\scalebox{\figscaletwo}{
\includegraphics{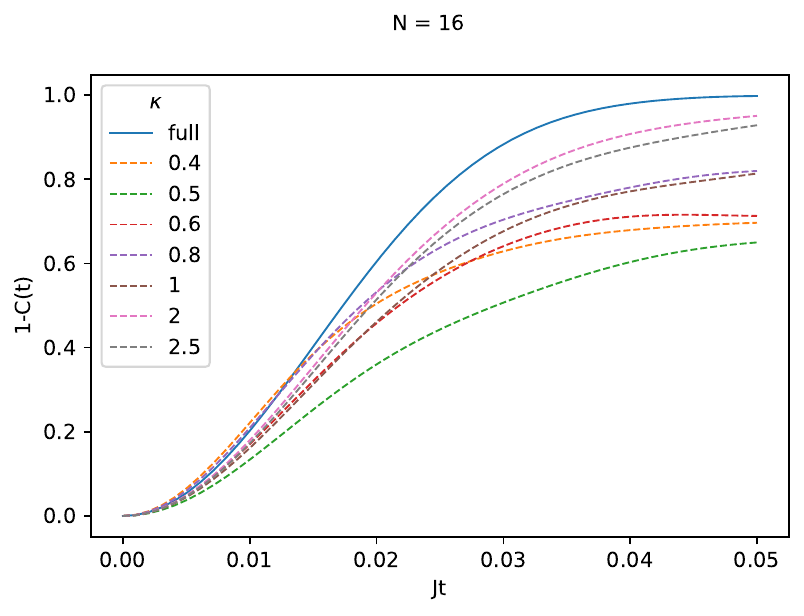}}
\scalebox{\figscaletwo}{
\includegraphics{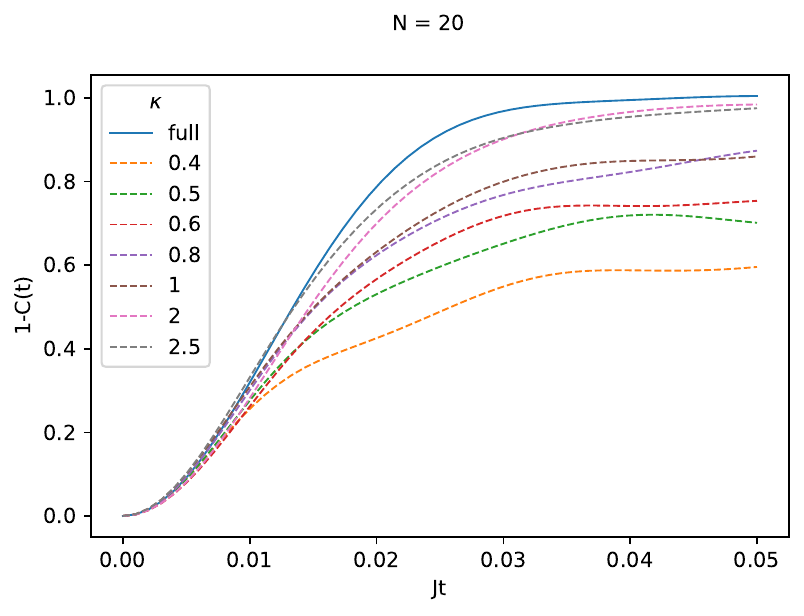}}
\end{center}
\caption{ Comparison of OTOC for different values of the sparsity parameter $\kappa$ with the full ensemble. 100 different Hamiltonians were used for each ensemble. Note that in order to compare the values for different $\kappa$, we need to rescale time by $1/\sqrt{p}$ with $p$ given in Eq.~\ref{Eq: sparsity parameter}. }\label{fig: Sparse Comparison}
\end{figure}

\cref{fig: Sparse Comparison} shows a comparison of the ensembles of the sparse model at different values of sparsity $\kappa$ with an ensemble of the full model. The early-time behavior of the sparse model is close to that of the full model at sparsity as low as $\kappa =2$. We expect a chaotic-to-non-chaotic transition as we lower the value of $\kappa$.       

\begin{figure}[htbp]
\begin{center}
\scalebox{\figscale}{
\includegraphics{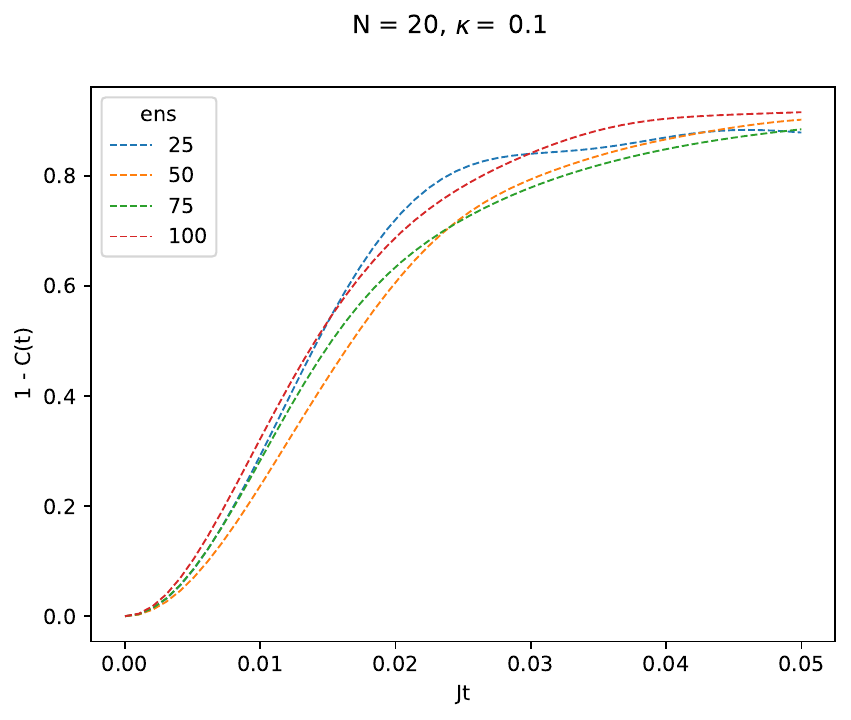}}
\scalebox{\figscale}{
\includegraphics{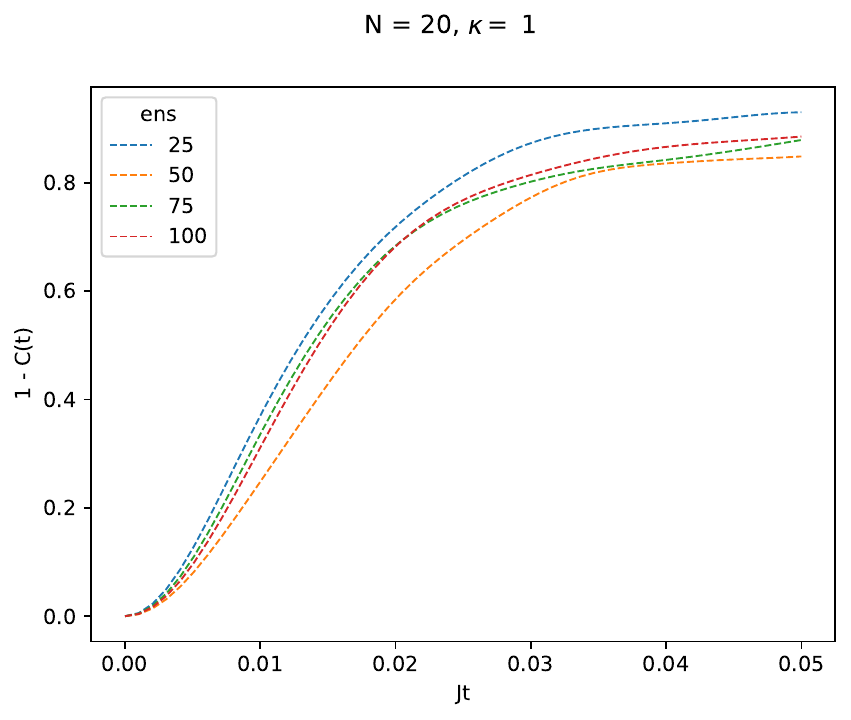}}
\scalebox{\figscale}{
\includegraphics{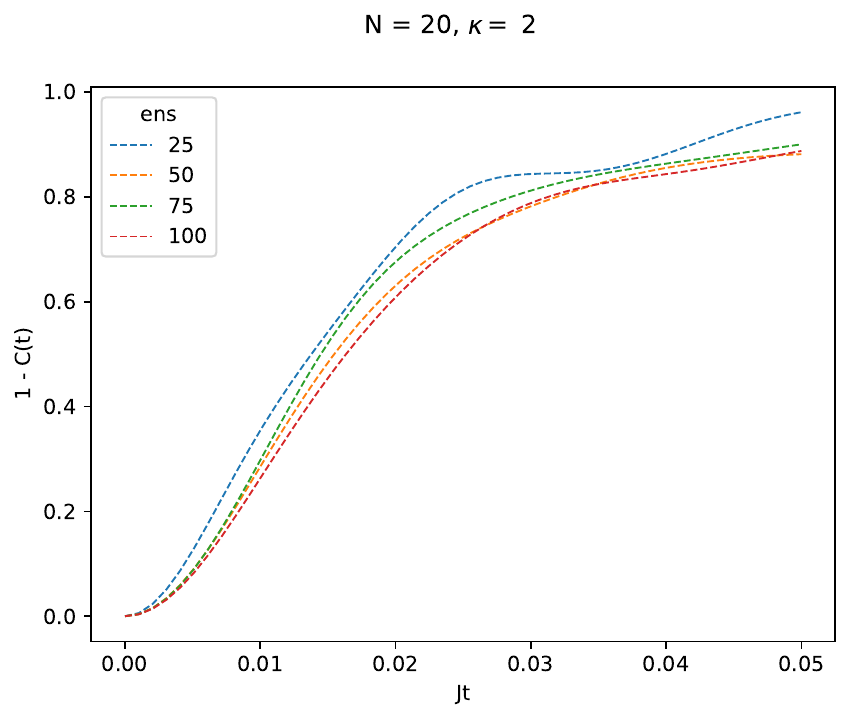}}
\end{center}
\caption{ Comparison of OTOC for different numbers of samples in an ensemble of the sparse bosonic SYK for $N=20$ and $\kappa = 0.1,1,2$.  }\label{fig: Sparse Ens Comparison}
\end{figure}

We also study the dependence of OTOC values on the number of samples in the ensemble for the sparse bosonic SYK model for different values of the sparsity. We report the results in \cref{fig: Sparse Ens Comparison} for $N=20$ and different $\kappa$. Unlike the full bosonic SYK model, where some degree of self-averaging is evident for $N=20$, the sparse model exhibits substantial variation in the OTOC across different numbers of Hamiltonian realizations. This indicates that the sparse model is 
non-self-averaging, at least up to the ensemble sizes accessible in our simulations. However, due to limitations on the computation time on the quantum hardware, we restrict the number of Hamiltonians in an ensemble for a quantum simulation to 30.

\begin{figure}[htbp]
\begin{center}
\scalebox{\figscaletwo}{
\includegraphics{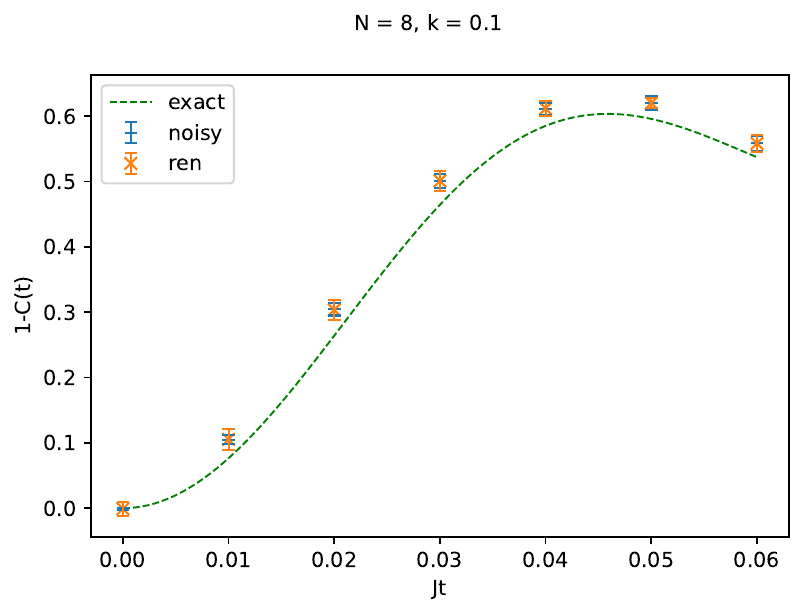}}
\scalebox{\figscaletwo}{
\includegraphics{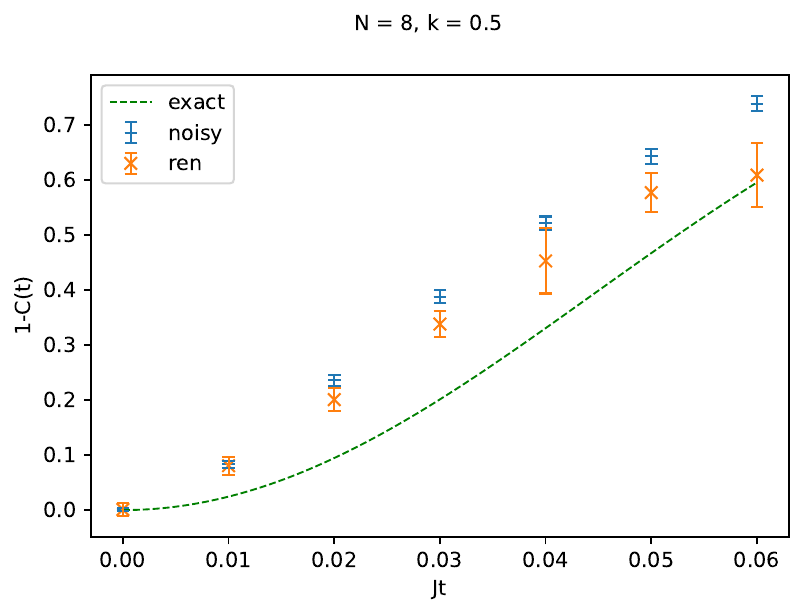}}
\scalebox{\figscaletwo}{
\includegraphics{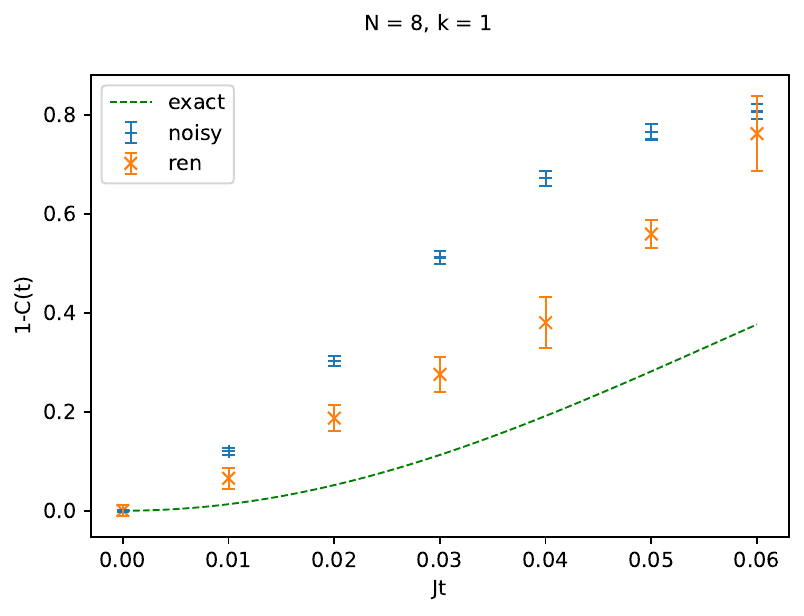}}
\scalebox{\figscaletwo}{
\includegraphics{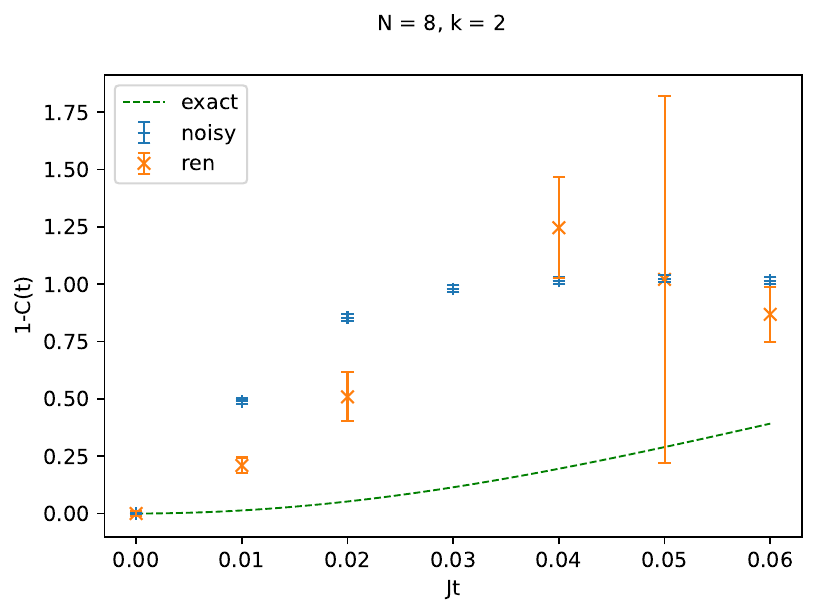}}
\end{center}
\caption{OTOC for $N = 8$ with $\kappa = 0.1,0.5,1,2$ respectively. For $\kappa =2$, we have removed the $Jt=0.03$ data point, which falls significantly outside the relevant range and adversely affects the clarity of the plotted results. The circuit depths are listed in Table \ref{Table:Depth for sims}.}\label{fig: N=8}
\end{figure}

Following the analysis for the sparsity of the Bosonic SYK model, we restrict our quantum simulations to $\kappa = 0.1,0.5,1,2$ for $N = 8,24$ and use an ensemble of 30 Hamiltonians to compute $C(t)$. The results for $N = 8$ are shown in \cref{fig: N=8}. In superconducting architectures where two-qubit operations are limited to nearest-neighbor qubits, the dominant contribution to the total two-qubit gate count arises from SWAP operations (see Table \ref{Table:Depth for sims} for two-qubit circuit depths). For small $N$ and $\kappa$, the two-qubit circuit depth remains low, and the noisy OTOC values stay close to their ideal counterparts, aside from a modest bias that can be addressed with appropriate error-mitigation techniques. In this shallow-circuit regime, $C^{\text{noisy}}_{V,\mathds{1}}(t)$ also remains near its ideal value of 1. As $\kappa$ increases, the two-qubit circuit depth grows, leading to a correspondingly larger accumulation of noise, as evident from the results for $\kappa = 0.5,1$. Although the observed values deviate substantially from the ideal behavior, the deviation appears to be dominated by an approximately constant offset factor that can, in principle, be corrected through mitigation. The quantity $C^{\text{noisy}}_{V,\mathds{1}}(t)$ likewise departs noticeably from its ideal value of 1, enabling it to serve as a normalization factor for the noisy OTOC. 

\begin{table}[htbp]
   \centering
    \begin{tabular}{|c|c||*{6}{c|} }
    \hline
     $N$ &\diagbox{$\kappa$}{Tr. step} & 1 & 2 & 3 & 4 & 5 & 6\\ 
     \hline
     8 & 0.1 & 8 & 16 & 24 & 32 & 40 & 48\\
       & 0.5 & 22 & 60 & 95 & 131 & 166 & 201 \\
       & 1 & 53 & 123 & 193 & 262 & 333 & 404\\
       & 2 & 117 & 258 & 399 & 539 & 680 & 821 \\
     \hline
     24 & 0.1 & 24 & 66 & 115 & 157 & 202 & 241\\
        & 0.5 & 152 & 371 & 584 & 807 & 1023 & 1223\\
        & 1 & 413 & 897 & 1356 & 1845 & 2371 & 2826\\
        & 2 & 832 & 1738 & 2574 & 3364 & 4072 & 5024\\
     \hline
    \end{tabular}
    \caption{
   Average two-qubit circuit depth for all Trotter steps for the simulations performed for $N = 8,24$ and various $\kappa$. The depths reported are post transpilation.
    }
    \label{Table:Depth for sims}
\end{table}

As the sparsity is increased further, the sparse model begins to display strongly chaotic behavior comparable to that of the full bosonic SYK model. The circuit depth grows substantially, particularly at later times, causing the noisy OTOC to approach zero. $C^{\text{noisy}}_{V,\mathds{1}}(t)$ is significantly affected by noise and approaches zero too, making the value of the renormalized OTOC (Eq.\ref{Eq: renormalized OTOC}) unstable and unsuitable\footnote{This was pointed out in the original paper \cite{Swingle:2018xvb} as well.}.

\begin{figure}[htbp]
\begin{center}
\scalebox{\figscaletwo}{
\includegraphics{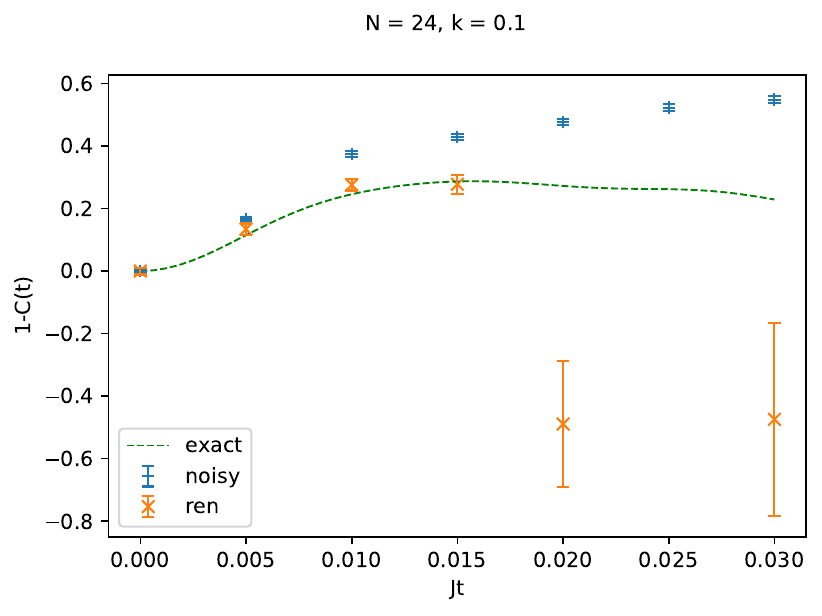}}
\scalebox{\figscaletwo}{
\includegraphics{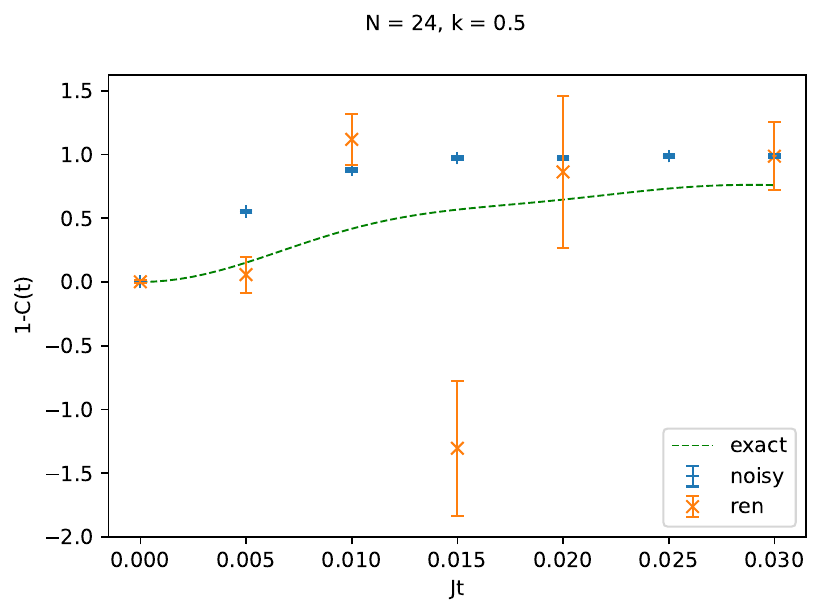}}
\scalebox{\figscaletwo}{
\includegraphics{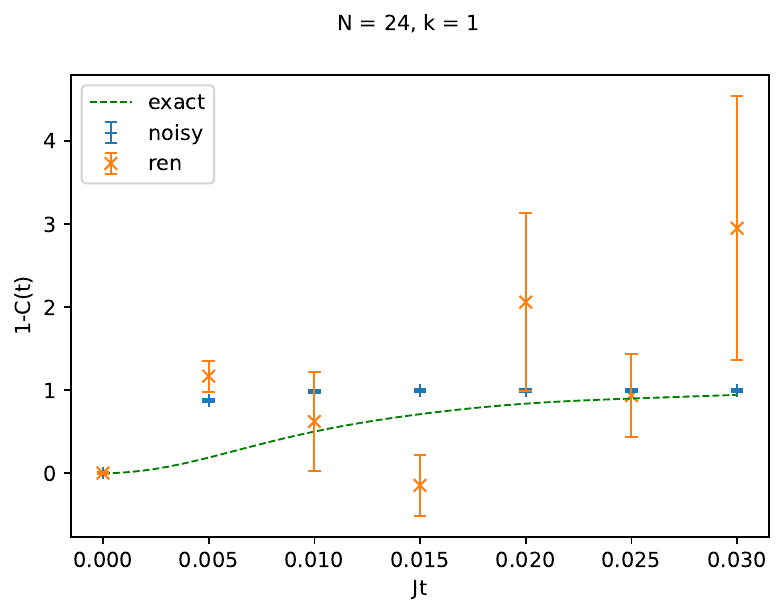}}
\scalebox{\figscaletwo}{
\includegraphics{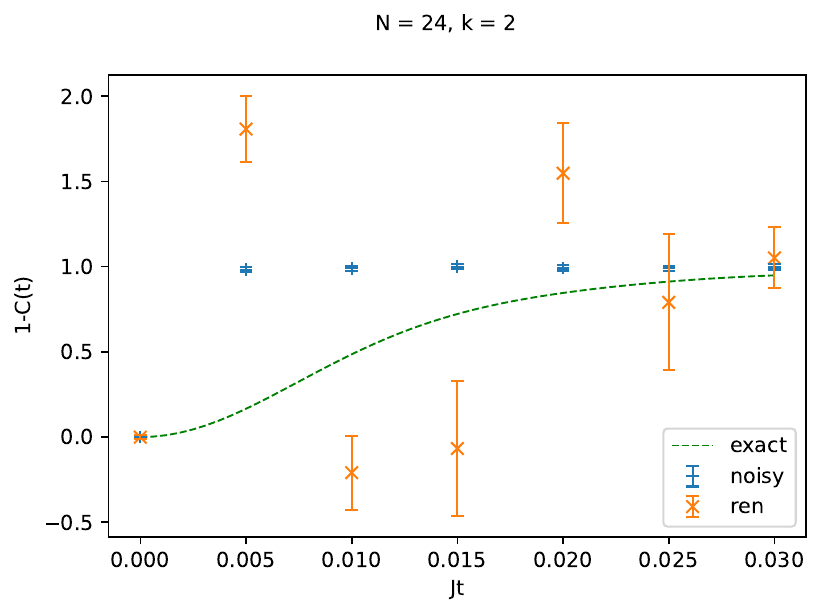}}
\end{center}
\caption{ OTOC for $N = 24$ with $\kappa = 0.1,0.5,1,2$ respectively. We have removed some data points for $\kappa = 0.1$ and $0.5$, as they fall significantly outside the relevant range and adversely affect the clarity of the plotted results. The circuit depths are listed in Table \ref{Table:Depth for sims}. }\label{fig: N=24}
\end{figure}

\begin{figure}[htbp]
\begin{center}
\scalebox{\figscaletwo}{
\includegraphics{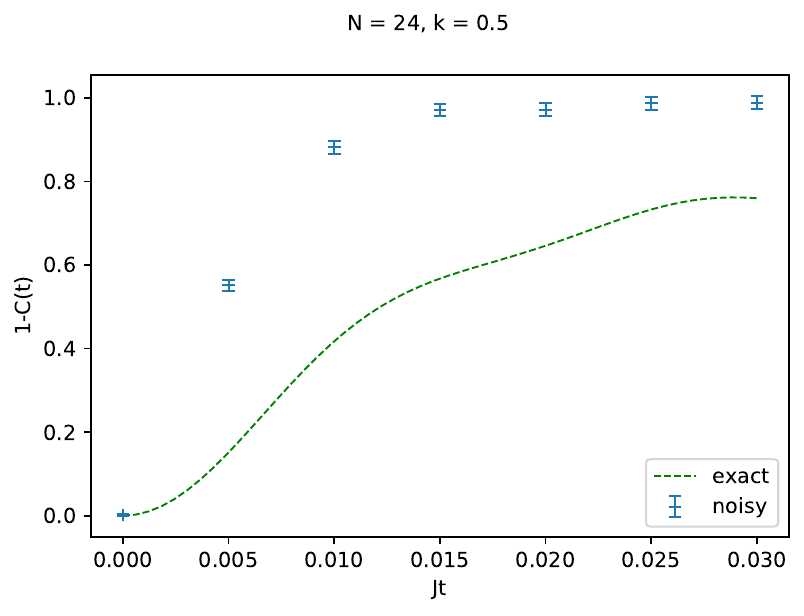}}
\scalebox{\figscaletwo}{
\includegraphics{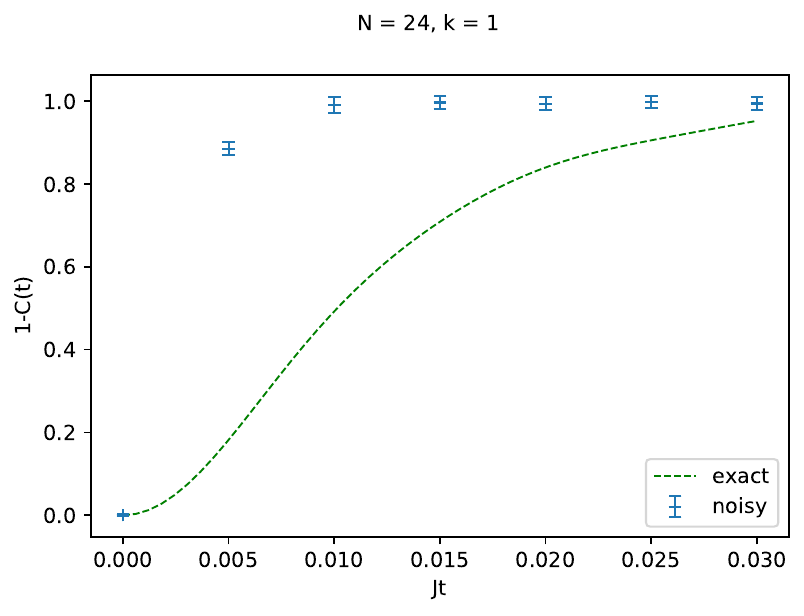}}
\scalebox{\figscaletwo}{
\includegraphics{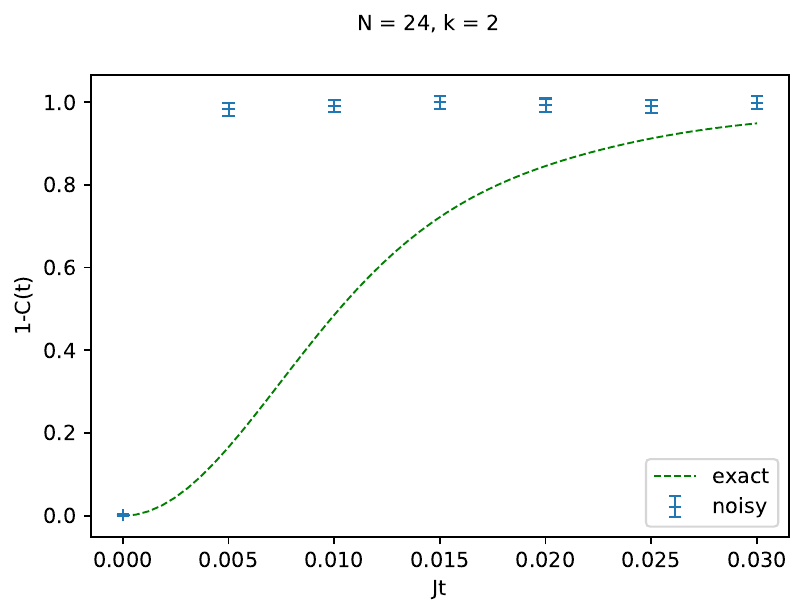}}
\end{center}
\caption{Noisy OTOC for $N = 24$ with $\kappa = 0.5,1,2$ respectively, without the renormalized values. }\label{fig: N=24 no_mit}
\end{figure}

To probe the chaotic behavior of the bosonic SYK model more reliably, simulations must be performed at larger values of $N$. In \cref{fig: N=24}, we present results for $N = 24$. The behavior as a function of $\kappa$ (and thus the corresponding two-qubit circuit depth) closely mirrors the trends observed for $N = 8$. For small $\kappa$ and early times, i.e. in the shallow-circuit regime,  the renormalization scheme yields values that closely track the ideal results. As the circuit depth increases, however, $C^{\text{noisy}}_{V,\mathds{1}}(t)$ approaches zero, rendering the renormalization procedure ineffective. Furthermore, as shown in \cref{fig: N=24 no_mit}, increasing chaoticity leads the noisy OTOC at the first Trotter step to approach zero, indicating near-complete scrambling of information. It would be interesting to ascertain whether the scrambling is due to the highly chaotic nature of the system under consideration or merely an artifact of the large circuit depth.

\section{Lessons for Quantum Advantage and Holography}
\label{sec:conclusion}

In this paper, we proposed the sparsified bosonic SYK model as a promising arena for exploring quantum advantage. We initiated a study of the model with classical simulators and quantum hardware, and discussed 
transitions between chaotic and non-chaotic behaviors as we change the sparsification probability $p$ (i.e.\ the sparsification parameter $\kappa$).

Our finding suggests that the system is strongly affected by noise in the present-day quantum hardware, particularly when we scale the problem to larger instances with deeper circuits. While we leave detailed analysis of the error mitigation for future work, we believe that this already demonstrates fundamental subtleties in the exploration of quantum advantage and the need to find a sweet spot in classical difficulty and efficient quantum encoding.
That we encounter difficulties in our quantum simulation may well have deeper reasons.

First, the OTOC is defined by backward propagation in time, and hence involves delicate cancellation between backward and forward time evolutions. Second, the system is very chaotic, and any small effect from noise could quickly result in a dramatic effect. Together, these points highlight the importance of carefully reducing the impact of errors when exploring near-term quantum advantage, which we leave to future work. Third, the system is defined not by a particular Hamiltonian, but rather an ensemble average of multiple Hamiltonians, and hence in principle requires many simulation runs a priori. One possibility is that the features of the system that make our system attractive for quantum advantage may well make the system sensitive to noise, and thereby require a more detailed evaluation of error mitigation and suppression methods, such as Pauli twirling \cite{Wallman:2015uzh}, dynamical decoupling \cite{Viola:1998jx}, and probabilistic error cancellation \cite{Berg:2022ugn,Temme:2016vkz}, which provide provable confidence bounds, which is essential for validated quantum advantage \cite{Lanes:2025kzh}.

Despite these subtleties, we should still be cautiously optimistic about the future prospects of demonstrating quantum advantage in these models. For example, while we have focused on specific sparsifications at the Hamiltonian level, one can 
attempt more hardware-specific sparsifications of the models (such as tailoring the model to the heavy-hex architecture of superconducting qubits), thereby further simplifying the model.
Our results call for a more systematic and detailed study of chaos/non-chaos transitions 
in sparsified bosonic SYK models, and more generally, models relevant for holography, as the model parameters on the ``edge of quantum chaos'' will be an ideal setup for exploring quantum advantage.
Of course, any serious claim of quantum advantage will necessarily require a dedicated study of 
classical simulation results, e.g.\ with the state-of-the-art tensor network methods.
It is tantalizing to imagine the connection between quantum advantage and the physics of quantum chaos and holography.
We hope to report our further results on this exciting direction in the near future.

\section*{Acknowledgements}

We thank Nate Earnest-Noble for stimulating discussions.
This research was supported primarily by the IBM-UTokyo laboratory.
The research of MY was also supported in part by the JSPS Grant-in-Aid for Scientific Research (No.\ 23K17689, 23K25865),  by JST, Japan (PRESTO Grant No.\ JPMJPR225A), and Moonshot R\&D Grant No.\ JPMJMS2061).

\bibliographystyle{ytphys}
\bibliography{refs}

@article{Swingle:2023nvv,
    author = "Swingle, Brian and Winer, Mike",
    title = "{Bosonic model of quantum holography}",
    eprint = "2311.01516",
    archivePrefix = "arXiv",
    primaryClass = "hep-th",
    doi = "10.1103/PhysRevB.109.094206",
    journal = "Phys. Rev. B",
    volume = "109",
    number = "9",
    pages = "094206",
    year = "2024"
}

@article{Berg:2022ugn,
    author = "Berg, Ewout van den and Minev, Zlatko K. and Kandala, Abhinav and Temme, Kristan",
    title = "{Probabilistic error cancellation with sparse Pauli{\textendash}Lindblad models on noisy quantum processors}",
    eprint = "2201.09866",
    archivePrefix = "arXiv",
    primaryClass = "quant-ph",
    doi = "10.1038/s41567-023-02042-2",
    journal = "Nature Phys.",
    volume = "19",
    number = "8",
    pages = "1116--1121",
    year = "2023"
}

@article{Wallman:2015uzh,
    author = "Wallman, Joel J. and Emerson, Joseph",
    title = "{Noise tailoring for scalable quantum computation via randomized compiling}",
    eprint = "1512.01098",
    archivePrefix = "arXiv",
    primaryClass = "quant-ph",
    doi = "10.1103/PhysRevA.94.052325",
    journal = "Phys. Rev. A",
    volume = "94",
    number = "5",
    pages = "052325",
    year = "2016"
}

@article{Viola:1998jx,
    author = "Viola, Lorenza and Lloyd, Seth",
    title = "{Dynamical suppression of decoherence in two state quantum systems}",
    eprint = "quant-ph/9803057",
    archivePrefix = "arXiv",
    doi = "10.1103/PhysRevA.58.2733",
    journal = "Phys. Rev. A",
    volume = "58",
    pages = "2733",
    year = "1998"
}

@article{Temme:2016vkz,
    author = "Temme, Kristan and Bravyi, Sergey and Gambetta, Jay M.",
    title = "{Error Mitigation for Short-Depth Quantum Circuits}",
    eprint = "1612.02058",
    archivePrefix = "arXiv",
    primaryClass = "quant-ph",
    doi = "10.1103/physrevlett.119.180509",
    journal = "Phys. Rev. Lett.",
    volume = "119",
    number = "18",
    pages = "180509",
    year = "2017"
}

@article{Hanada:2023rkf,
    author = "Hanada, Masanori and Jevicki, Antal and Liu, Xianlong and Rinaldi, Enrico and Tezuka, Masaki",
    title = "{A model of randomly-coupled Pauli spins}",
    eprint = "2309.15349",
    archivePrefix = "arXiv",
    primaryClass = "hep-th",
    reportNumber = "RIKEN-iTHEMS-Report-23",
    doi = "10.1007/JHEP05(2024)280",
    journal = "JHEP",
    volume = "05",
    pages = "280",
    year = "2024"
}

@article{Xu:2020shn,
    author = "Xu, Shenglong and Susskind, Leonard and Su, Yuan and Swingle, Brian",
    title = "{A Sparse Model of Quantum Holography}",
    eprint = "2008.02303",
    archivePrefix = "arXiv",
    primaryClass = "cond-mat.str-el",
    month = "8",
    year = "2020"
}

@article{Swingle:2016var,
    author = "Swingle, Brian and Bentsen, Gregory and Schleier-Smith, Monika and Hayden, Patrick",
    title = "{Measuring the scrambling of quantum information}",
    eprint = "1602.06271",
    archivePrefix = "arXiv",
    primaryClass = "quant-ph",
    doi = "10.1103/PhysRevA.94.040302",
    journal = "Phys. Rev. A",
    volume = "94",
    number = "4",
    pages = "040302",
    year = "2016"
}

@article{Swingle:2018xvb,
    author = "Swingle, Brian and Yunger Halpern, Nicole",
    title = "{Resilience of scrambling measurements}",
    eprint = "1802.01587",
    archivePrefix = "arXiv",
    primaryClass = "quant-ph",
    reportNumber = "NSF-ITP-18-003, NSF-ITP-18-003",
    doi = "10.1103/PhysRevA.97.062113",
    journal = "Phys. Rev. A",
    volume = "97",
    number = "6",
    pages = "062113",
    year = "2018"
}

@article{Sachdev:1992fk,
    author = "Sachdev, Subir and Ye, Jinwu",
    title = "{Gapless spin fluid ground state in a random, quantum Heisenberg magnet}",
    eprint = "cond-mat/9212030",
    archivePrefix = "arXiv",
    reportNumber = "PRINT-93-0077",
    doi = "10.1103/PhysRevLett.70.3339",
    journal = "Phys. Rev. Lett.",
    volume = "70",
    pages = "3339",
    year = "1993"
}

@article{Maldacena:2016hyu,
    author = "Maldacena, Juan and Stanford, Douglas",
    title = "{Remarks on the Sachdev-Ye-Kitaev model}",
    eprint = "1604.07818",
    archivePrefix = "arXiv",
    primaryClass = "hep-th",
    doi = "10.1103/PhysRevD.94.106002",
    journal = "Phys. Rev. D",
    volume = "94",
    number = "10",
    pages = "106002",
    year = "2016"
}

@article{Maldacena:2015waa,
    author = "Maldacena, Juan and Shenker, Stephen H. and Stanford, Douglas",
    title = "{A bound on chaos}",
    eprint = "1503.01409",
    archivePrefix = "arXiv",
    primaryClass = "hep-th",
    doi = "10.1007/JHEP08(2016)106",
    journal = "JHEP",
    volume = "08",
    pages = "106",
    year = "2016"
}

@article{Feynman:1981tf,
    author = "Feynman, Richard P.",
    editor = "Brown, L. M.",
    title = "{Simulating physics with computers}",
    doi = "10.1007/BF02650179",
    journal = "Int. J. Theor. Phys.",
    volume = "21",
    pages = "467--488",
    year = "1982"
}

@article{Shenker:2013pqa,
    author = "Shenker, Stephen H. and Stanford, Douglas",
    title = "{Black holes and the butterfly effect}",
    eprint = "1306.0622",
    archivePrefix = "arXiv",
    primaryClass = "hep-th",
    reportNumber = "SU-ITP-13-08",
    doi = "10.1007/JHEP03(2014)067",
    journal = "JHEP",
    volume = "03",
    pages = "067",
    year = "2014"
}

@article{Abanin:2025rbz,
    author = "Abanin, Dmitry A. and others",
    title = "{Constructive interference at the edge of quantum ergodic dynamics}",
    eprint = "2506.10191",
    archivePrefix = "arXiv",
    primaryClass = "quant-ph",
    month = "6",
    year = "2025"
}

@misc{Kitaev:2015,
    author = "Alexei Kitaev",
    title = "{A simple model of quantum holography}",
    year = "2015",
    note = {\url{https://online.kitp.ucsb.edu/online/entangled15/kitaev/}}
}

@article{larkin1969quasiclassical,
  title={Quasiclassical method in the theory of superconductivity},
  author={Larkin, Anatoly I and Ovchinnikov, Yu N},
  journal={Sov Phys JETP},
  volume={28},
  number={6},
  pages={1200--1205},
  year={1969}
}

@article{aleiner1997role,
  title={Role of divergence of classical trajectories in quantum chaos},
  author={Aleiner, IL and Larkin, AI},
  journal={Physical Review E},
  volume={55},
  number={2},
  pages={R1243},
  year={1997},
  publisher={APS}
}

@article{aleiner1996divergence,
  title={Divergence of classical trajectories and weak localization},
  author={Aleiner, IL and Larkin, AI},
  journal={Physical Review B},
  volume={54},
  number={20},
  pages={14423},
  year={1996},
  publisher={APS}
}

@article{agam2000shot,
  title={Shot noise in chaotic systems:“classical” to quantum crossover},
  author={Agam, Oded and Aleiner, Igor and Larkin, Anatoly},
  journal={Physical review letters},
  volume={85},
  number={15},
  pages={3153},
  year={2000},
  publisher={APS}
}

@article{luo2019quantum,
  title={Quantum simulation of the non-fermi-liquid state of Sachdev-Ye-Kitaev model},
  author={Luo, Zhihuang and You, Yi-Zhuang and Li, Jun and Jian, Chao-Ming and Lu, Dawei and Xu, Cenke and Zeng, Bei and Laflamme, Raymond},
  journal={npj Quantum Information},
  volume={5},
  number={1},
  pages={53},
  year={2019},
  publisher={Nature Publishing Group UK London}
}

@article{granet2025simulating,
  title={Simulating sparse SYK model with a randomized algorithm on a trapped-ion quantum computer},
  author={Granet, Etienne and Kikuchi, Yuta and Dreyer, Henrik and Rinaldi, Enrico},
  journal={arXiv preprint arXiv:2507.07530},
  year={2025}
}

@inproceedings{Baumgartner:2024ysk,
    author = "Baumgartner, Rahel and Pelliconi, Pietro and Bandyopadhyay, Soumik and Orsi, Francesca and Sauerwein, Nick and Hauke, Philipp and Brantut, Jean-Philippe and Sonner, Julian",
    title = "{Quantum simulation of the Sachdev-Ye-Kitaev model using time-dependent disorder in optical cavities}",
    eprint = "2411.17802",
    archivePrefix = "arXiv",
    primaryClass = "quant-ph",
    month = "11",
    year = "2024"
}

@article{asaduzzaman2024sachdev,
  title={Sachdev-Ye-Kitaev model on a noisy quantum computer},
  author={Asaduzzaman, Muhammad and Jha, Raghav G and Sambasivam, Bharath},
  journal={Physical Review D},
  volume={109},
  number={10},
  pages={105002},
  year={2024},
  publisher={APS}
}

@article{Tezuka:2022mrr,
    author = "Tezuka, Masaki and Oktay, Onur and Rinaldi, Enrico and Hanada, Masanori and Nori, Franco",
    title = "{Binary-coupling sparse Sachdev-Ye-Kitaev model: An improved model of quantum chaos and holography}",
    eprint = "2208.12098",
    archivePrefix = "arXiv",
    primaryClass = "quant-ph",
    reportNumber = "DMUS-MP-22/17, RIKEN-iTHEMS-Report-22",
    doi = "10.1103/PhysRevB.107.L081103",
    journal = "Phys. Rev. B",
    volume = "107",
    number = "8",
    pages = "L081103",
    year = "2023"
}

@article{Garcia-Garcia:2020cdo,
    author = "Garc{\'\i}a-Garc{\'\i}a, Antonio M. and Jia, Yiyang and Rosa, Dario and Verbaarschot, Jacobus J. M.",
    title = "{Sparse Sachdev-Ye-Kitaev model, quantum chaos and gravity duals}",
    eprint = "2007.13837",
    archivePrefix = "arXiv",
    primaryClass = "hep-th",
    doi = "10.1103/PhysRevD.103.106002",
    journal = "Phys. Rev. D",
    volume = "103",
    number = "10",
    pages = "106002",
    year = "2021"
}

@article{Orman:2024mpw,
    author = "Orman, Patrick and Gharibyan, Hrant and Preskill, John",
    title = "{Quantum chaos in the sparse SYK model}",
    eprint = "2403.13884",
    archivePrefix = "arXiv",
    primaryClass = "hep-th",
    doi = "10.1007/JHEP02(2025)173",
    journal = "JHEP",
    volume = "02",
    pages = "173",
    year = "2025"
}

@article{Li:2016xhw,
    author = "Li, Jun and Fan, Ruihua and Wang, Hengyan and Ye, Bingtian and Zeng, Bei and Zhai, Hui and Peng, Xinhua and Du, Jiangfeng",
    title = "{Measuring Out-of-Time-Order Correlators on a Nuclear Magnetic Resonance Quantum Simulator}",
    eprint = "1609.01246",
    archivePrefix = "arXiv",
    primaryClass = "cond-mat.str-el",
    doi = "10.1103/PhysRevX.7.031011",
    journal = "Phys. Rev. X",
    volume = "7",
    number = "3",
    pages = "031011",
    year = "2017"
}

@article{Garttner:2016mqj,
    author = {G{\"a}rttner, Martin and Bohnet, Justin G. and Safavi-Naini, Arghavan and Wall, Michael L. and Bollinger, John J. and Rey, Ana Maria},
    title = "{Measuring out-of-time-order correlations and multiple quantum spectra in a trapped ion quantum magnet}",
    eprint = "1608.08938",
    archivePrefix = "arXiv",
    primaryClass = "quant-ph",
    doi = "10.1038/nphys4119",
    journal = "Nature Phys.",
    volume = "13",
    pages = "781",
    year = "2017"
}

@article{Landsman:2018jpm,
    author = "Landsman, Kevin A. and Figgatt, Caroline and Schuster, Thomas and Linke, Norbert M. and Yoshida, Beni and Yao, Norman Y. and Monroe, Christopher",
    title = "{Verified Quantum Information Scrambling}",
    eprint = "1806.02807",
    archivePrefix = "arXiv",
    primaryClass = "quant-ph",
    doi = "10.1038/s41586-019-0952-6",
    journal = "Nature",
    volume = "567",
    number = "7746",
    pages = "61--65",
    year = "2019"
}

@article{Blok:2020may,
    author = "Blok, M. S. and Ramasesh, V. V. and Schuster, T. and O'Brien, K. and Kreikebaum, J. M. and Dahlen, D. and Morvan, A. and Yoshida, B. and Yao, N. Y. and Siddiqi, I.",
    title = "{Quantum Information Scrambling on a Superconducting Qutrit Processor}",
    eprint = "2003.03307",
    archivePrefix = "arXiv",
    primaryClass = "quant-ph",
    doi = "10.1103/PhysRevX.11.021010",
    journal = "Phys. Rev. X",
    volume = "11",
    number = "2",
    pages = "021010",
    year = "2021"
}

@article{Mi:2021gdf,
    author = "Mi, Xiao and others",
    title = "{Information scrambling in quantum circuits}",
    eprint = "2101.08870",
    archivePrefix = "arXiv",
    primaryClass = "quant-ph",
    doi = "10.1126/science.abg5029",
    journal = "Science",
    volume = "374",
    number = "6574",
    pages = "abg5029",
    year = "2021"
}

@article{Braumuller:2021cic,
    author = {Braum{\"u}ller, Jochen and others},
    title = "{Probing quantum information propagation with out-of-time-ordered correlators}",
    eprint = "2102.11751",
    archivePrefix = "arXiv",
    primaryClass = "quant-ph",
    doi = "10.1038/s41567-021-01430-w",
    journal = "Nature Phys.",
    volume = "18",
    number = "2",
    pages = "172--178",
    year = "2022"
}

@article{Geller:2021kpx,
    author = {Geller, Michael R. and Arrasmith, Andrew and Holmes, Zo{\"e} and Yan, Bin and Coles, Patrick J. and Sornborger, Andrew},
    title = "{Quantum simulation of operator spreading in the chaotic Ising model}",
    eprint = "2106.16170",
    archivePrefix = "arXiv",
    primaryClass = "quant-ph",
    doi = "10.1103/PhysRevE.105.035302",
    journal = "Phys. Rev. E",
    volume = "105",
    number = "3",
    pages = "035302",
    year = "2022"
}

@article{Vermersch:2018sru,
    author = "Vermersch, Benoit and Elben, Andreas and Sieberer, Lukas M. and Yao, Norman Y. and Zoller, Peter",
    title = "{Probing scrambling using statistical correlations between randomized measurements}",
    eprint = "1807.09087",
    archivePrefix = "arXiv",
    primaryClass = "quant-ph",
    doi = "10.1103/PhysRevX.9.021061",
    journal = "Phys. Rev. X",
    volume = "9",
    number = "2",
    pages = "021061",
    year = "2019"
}

@article{Seki:2024rfx,
    author = "Seki, Kazuhiro and Kikuchi, Yuta and Hayata, Tomoya and Yunoki, Seiji",
    title = "{Simulating Floquet scrambling circuits on trapped-ion quantum computers}",
    eprint = "2405.07613",
    archivePrefix = "arXiv",
    primaryClass = "quant-ph",
    reportNumber = "RIKEN-iTHEMS-Report-24",
    doi = "10.1103/PhysRevResearch.7.023032",
    journal = "Phys. Rev. Res.",
    volume = "7",
    number = "2",
    pages = "023032",
    year = "2025"
}

@article{Lanes:2025kzh,
    author = "Lanes, Olivia and others",
    title = "{A Framework for Quantum Advantage}",
    eprint = "2506.20658",
    archivePrefix = "arXiv",
    primaryClass = "quant-ph",
    month = "6",
    year = "2025"
}

@article{Shor:1994jg,
    author = "Shor, Peter W.",
    title = "{Polynomial time algorithms for prime factorization and discrete logarithms on a quantum computer}",
    eprint = "quant-ph/9508027",
    archivePrefix = "arXiv",
    doi = "10.1137/S0097539795293172",
    journal = "SIAM J. Sci. Statist. Comput.",
    volume = "26",
    pages = "1484",
    year = "1997"
}

@article{Deutsch:1985vkq,
    author = "Deutsch, David",
    title = "{Quantum theory, the Church{\textendash}Turing principle and the universal quantum computer | Proceedings of the Royal Society of London. A. Mathematical and Physical Sciences}",
    doi = "10.1098/rspa.1985.0070",
    journal = "Proc. Roy. Soc. Lond. A",
    volume = "Volume 400",
    number = "Issue 1818",
    pages = "97--117",
    year = "1985"
}

@article{Preskill:2018jim,
    author = "Preskill, John",
    title = "{Quantum Computing in the NISQ era and beyond}",
    eprint = "1801.00862",
    archivePrefix = "arXiv",
    primaryClass = "quant-ph",
    doi = "10.22331/q-2018-08-06-79",
    journal = "Quantum",
    volume = "2",
    pages = "79",
    year = "2018"
}

\end{document}